\documentclass[12pt,journal,onecolumn]{IEEEtran}
\usepackage{cite}
\usepackage{amsmath,amssymb,amsfonts}
\usepackage{algorithmic}
\usepackage{graphicx}
\usepackage{textcomp}

\usepackage{booktabs}

\def\BibTeX{{\rm B\kern-.05em{\sc i\kern-.025em b}\kern-.08em
    T\kern-.1667em\lower.7ex\hbox{E}\kern-.125emX}}
\begin{document}

\title{Operational Memory Architecture for Kubernetes: Preserving
Causal Context Across the Evidence Horizon}

\author{Shamsher~Khan,~\IEEEmembership{Senior Member,~IEEE}\\
GlobalLogic (Hitachi Group), Tampa Bay Area, FL, USA\\
\textit{shamsher.khan.research@gmail.com}%

\thanks{The implementation, experimental data, and evaluation scripts
described in this paper are publicly available at
https://github.com/opscart/k8s-causal-memory.
e-mail: shamsher.khan.research@gmail.com)
Raw event logs, SQLite databases, query outputs, and automation scripts
for all experiments are committed under \texttt{docs/poc-results/} for
independent verification and reproduction.}%
\thanks{S. Khan is with GlobalLogic (Hitachi Group), Tampa Bay Area,
FL, USA. e-mail: shamsher.khan.research@gmail.com}}

\markboth{Khan: Operational Memory Architecture for Kubernetes}%
{Khan: Operational Memory Architecture for Kubernetes}
\maketitle 
\begin{abstract}
{\sloppy
Kubernetes clusters generate rich operational events during pod lifecycle
transitions, yet the platform's native event retention model systematically
discards the most diagnostically valuable context. The \texttt{LastTerminationState}
field, which records the reason and exit code of a container's most recent
failure, is overwritten within approximately 90 seconds of a pod restart---a
boundary we term the \emph{evidence horizon}, characterized experimentally
in Section~V. During high-frequency crash loops, this horizon can be crossed
multiple times before an engineer reaches a terminal, permanently destroying
forensic evidence.
}

This paper presents the Operational Memory Architecture (OMA), formalized
as a first-class architectural primitive for Kubernetes operational state
preservation. OMA explicitly encodes evidence rotation constraints and
causal preservation as design requirements, an aspect not addressed by
existing observability or logging models. OMA introduces three encoded
causal patterns: P001 (OOMKill causal chain), P002 (ConfigMap environment
variable silent misconfiguration), and P003 (ConfigMap volume mount symlink
swap propagation). We implement OMA as an open-source system comprising a
Go-based Kubernetes watcher, a SQLite operational memory store, and a
canonical three-query interface. We validate the architecture through
reproducible experiments on Minikube and Azure Kubernetes Service
(AKS~1.32.10), and conduct a 30-run statistical latency analysis and
concurrent stress evaluation with up to 20 simultaneous crash-looping pods.
Intra-cycle causal edges are constructed with a mean latency of 0.702\,ms
($\sigma=0.31$\,ms). The collector processes 2.86 events/sec under 20
concurrent OOMKill pods while consuming only 8.8\,MB RAM, confirming
linear scaling and minimal operational overhead. While validated on Minikube and Azure Kubernetes Service, OMA addresses
a fundamental property of the Kubernetes architecture itself: the
evidence horizon arises from the kubelet's state rotation semantics,
which are consistent across all conformant Kubernetes distributions
and managed services.
\end{abstract}


\section{Introduction}
\label{sec:introduction}

Modern cloud-native applications run as collections of containerized
microservices orchestrated by Kubernetes. As these deployments grow in scale
and complexity, the operational challenge of diagnosing failures has become a
significant engineering bottleneck. When a container crashes, the immediate
question is not merely that it crashed, but why: what configuration was active
at the moment of failure, what was the state of the node, and has this pattern
occurred before?

Kubernetes provides a rich event stream through its API server, but this stream
is ephemeral by design. The platform's native garbage collection removes events
after one hour by default, and---more critically---the \texttt{LastTerminationState}
field in a pod's \texttt{ContainerStatus} is overwritten the moment a container
restarts. This creates what we term the \emph{evidence horizon}: a narrow window,
experimentally characterized at approximately 90 seconds in Section~V, during
which post-mortem investigation is possible. Beyond this window, the direct
evidence of an OOMKill, including its exit code, resource consumption, and the
ConfigMap configuration in effect at the time, is permanently lost.

The consequences of this architectural gap are measurable in production
environments operating hundreds of cores across multiple Kubernetes clusters
under strict compliance requirements. When a memory-constrained pod enters a
crash loop, the on-call engineer faces a degraded diagnostic environment:
metrics show CPU and memory trends, logs show application output, but neither
preserves the exact resource limits active at kill time, the specific ConfigMap
values in effect, or the causal relationship between a preceding node pressure
event and the subsequent container termination. This information exists---briefly---in
Kubernetes state, but the evidence horizon passes before it can be preserved.

Existing observability solutions address adjacent problems. Prometheus~\cite{b1}
excels at metric time-series but has no concept of causal relationships between
discrete events. Jaeger~\cite{b2} and similar distributed tracing systems capture
request-level causality but are blind to infrastructure-level events.
Log aggregation platforms such as Elasticsearch~\cite{b3} preserve output but
not state. OpenTelemetry~\cite{b4} provides a unified instrumentation standard
but focuses on application-emitted telemetry rather than Kubernetes control
plane events. Critically, none of these tools can answer the question: what was
the exact configuration state of this pod at the moment it failed, and what
caused the failure? The absence of this capability is not a missing feature
of individual tools---it is a structural consequence of the evidence horizon
that no existing tool addresses at the architectural level.

\textbf{Contributions.} OMA formalizes operational memory as a first-class
architectural primitive in Kubernetes, explicitly encoding evidence rotation
constraints and causal preservation---an aspect not addressed by existing
observability or logging models. This paper makes the following specific
contributions:

\begin{enumerate}
\item We formally define the evidence horizon as a systems property of
Kubernetes and experimentally characterize its timing across two independent
cluster environments (Section~III).
\item We propose OMA, a four-layer framework positioning operational memory
as an architectural primitive alongside metrics, logs, and traces in the
Kubernetes observability stack (Section~IV).
\item We implement OMA as an open-source system and conduct a 30-run
statistical latency evaluation demonstrating intra-cycle causal edge
construction at mean 0.702\,ms, and a stress evaluation confirming linear
event scaling with flat memory consumption (Sections~V and~VI).
\item We define three canonical queries (Q1: causal chain, Q2: pattern
history, Q3: point-in-time state) and demonstrate their utility against
evidence that no existing tool can recover after the evidence horizon passes.
\end{enumerate}

\section{Background}
\label{sec:background}

\subsection{Kubernetes Event Model}

Kubernetes represents the cluster state as a collection of objects stored in
etcd~\cite{b5}. Events are first-class objects of kind \texttt{Event} that
reference other objects via \texttt{InvolvedObject}. The API server generates
events for pod scheduling, container state transitions, node conditions, and
control plane operations. By default, events are retained for one hour and
are not persisted across etcd compactions.

The Pod object's status subresource contains the per-container status in the
\texttt{ContainerStatus} array. Each entry includes \texttt{State} (current
state), \texttt{LastTerminationState} (state of the most recent termination),
and \texttt{RestartCount}. The \texttt{LastTerminationState.Terminated} field
includes \texttt{Reason}, \texttt{ExitCode}, \texttt{StartedAt}, and
\texttt{FinishedAt}-- precise forensic data for diagnosing the cause of the
most recent failure.

\subsection{The Evidence Horizon}

The evidence horizon arises from the interaction of three Kubernetes behaviors.
First, when a container restarts, the kubelet overwrites
\texttt{LastTerminationState} with data from the current termination cycle,
discarding the previous entry. Second, the kubelet's garbage collection policy
for terminated containers defaults to a maximum of one terminated container
per pod retained on the node. Third, Kubernetes Events referencing the
terminated container are subject to the one-hour retention policy but may be
de-duplicated or overwritten during high-frequency crash loops.

In practice, the evidence horizon for \texttt{LastTerminationState} is
approximately 90 seconds, the time between a container restart and the
subsequent restart that overwrites the previous termination state. This
value is derived experimentally in Section~V: During 30 independent runs
on Minikube and confirmed on AKS~1.32.10, we observe multiple restart cycles
within 90-second windows, with the first restart occurring at 15--30 seconds
and \texttt{LastTerminationState} overwritten by the second restart. During
a \texttt{CrashLoopBackOff} with exponential backoff, the window may extend
for later restart cycles, but for memory-constrained pods that restart rapidly,
the initial window can be as short as 15 seconds.

\subsection{ConfigMap Propagation Semantics}

Kubernetes ConfigMaps are consumed by pods in two distinct modes with
fundamentally different propagation semantics. When consumed as environment
variables through \texttt{envFrom} or \texttt{env.\allowbreak valueFrom.\allowbreak configMapKeyRef},
the values are resolved at container startup via the \texttt{execve(2)} system
call and baked into the process environment. Subsequent ConfigMap updates have
no effect on running containers -- the kernel provides no mechanism to update a
process's environment after fork/exec. This creates the silent misconfiguration
pattern: an operator updates a ConfigMap believing the change is live, while
running pods continue operating with the previous values indefinitely.

When consumed as volume mounts, the kubelet maintains a projected volume backed
by the ConfigMap content. On ConfigMap update, the kubelet performs an atomic
symlink swap on the \texttt{..data} directory within the projected volume,
replacing the entire directory contents atomically. The propagation delay
between a ConfigMap update and the completion of the kubelet symlink swap is
typically 10--90 seconds, depending on the kubelet's sync period configuration.

\section{Related Work}
\label{sec:related}

\subsection{Metrics-Based Observability}

Prometheus~\cite{b1} is the de facto standard for Kubernetes metrics collection,
providing powerful time-series query capabilities through PromQL with native
Kubernetes service discovery. However, Prometheus operates on numeric metrics
sampled at fixed intervals and does not have a native concept of discrete events or
causal relationships between them. Fundamentally, Prometheus cannot solve the
evidence horizon problem because it has no object snapshot model: it cannot
record the exact resource limits of a container at kill time, nor preserve
ConfigMap state at a specific past timestamp. A Prometheus alert can notify
that a pod's restart count increased; it cannot reconstruct what caused the
failure or what configuration was in effect.

\subsection{Distributed Tracing}

Jaeger~\cite{b2} implements distributed request tracing, capturing causality
at the application request level as requests propagate through microservices.
These systems are fundamentally application-instrumented and do not observe
Kubernetes infrastructure events. The causal chain between a node memory
pressure condition and a container OOMKill is invisible to distributed tracing
systems, because neither event originates from application instrumentation.
Distributed tracing cannot solve the evidence horizon problem because it
operates above the container runtime boundary.

\subsection{Log Aggregation}

Elasticsearch and related log aggregation platforms~\cite{b3} aggregate
application and system logs into searchable indices. Log-based root cause
analysis~\cite{b6} and anomaly detection approaches have been extensively explored. However, log aggregation cannot solve the evidence horizon problem
because it captures application output, not Kubernetes object state. The exact
memory limit of a container at kill time, or the specific ConfigMap revision
in effect, is not present in application logs unless the application explicitly
logs its own configuration---a practice not universally followed and insufficient
for post-mortem analysis of infrastructure-level failures.

\subsection{Kubernetes-Native Tools}

The \texttt{kubectl} command-line tool~\cite{b5} provides direct access to
the Kubernetes API state but is inherently present-tense: it describes the current
state of objects, not historical states. Once a pod is deleted or its
\texttt{ContainerStatus} is overwritten, the previous state is inaccessible.
The Kubernetes audit log~\cite{b7} captures API server requests and can
reconstruct object state changes, but requires cluster-level configuration,
generates high log volume, and is not designed for causal query patterns.
Neither tool addresses the evidence horizon as a design requirement.

\subsection{Causal Inference in Distributed Systems}

Causal inference in distributed systems has been extensively studied~\cite{b8}.
Vector clocks and occurred-before relationships~\cite{b9} provide formal
foundations for establishing causal ordering in distributed logs. Recent work
on causal anomaly detection~\cite{b10} applies causal graph models to the diagnosis of
microservice failure. OMA differs from these approaches in a
fundamental way: it focuses on infrastructure-level causal patterns specific
to the Kubernetes operational model, encoding domain knowledge about evidence
rotation as a first-class architectural concern. The causal edges OMA constructs
are analogous to happened-before relationships~\cite{b9}: if an OOMKillEvidence
event $e_2$ is observed for pod $P$ within 90 seconds of an OOMKill event $e_1$
for the same pod, then $e_1 \rightarrow e_2$ in the happened-before sense.

\section{Operational Memory Architecture}
\label{sec:architecture}

\subsection{Design Principles}

OMA is designed around three principles. First, \emph{evidence preservation
before rotation}: the system must capture relevant context within the evidence
horizon, not after. Second, \emph{operational causality construction}: OMA
does not attempt statistical causal inference in the Pearlian sense~\cite{b8};
instead, it constructs operational causality using domain-specific temporal
constraints and Kubernetes semantics, analogous to happened-before relationships
in distributed systems~\cite{b9}. This distinction is important: OMA's causal
edges encode what an experienced SRE would recognize as a causal relationship
based on domain knowledge, not a statistically inferred dependency. Third,
\emph{query-first design}: the system is optimized for three specific query
patterns that address real operational questions.

\subsection{Four-Layer Architecture}

The OMA comprises four layers, each with a distinct responsibility, as illustrated
in Fig.~\ref{fig2}.

\textit{Layer~1---Event Collection.} A Go-based Kubernetes watcher connects to
the cluster API server via the \texttt{client-go} informer framework and
subscribes to watch streams for Pod, Node, and ConfigMap objects. The watcher
emits structured \texttt{CausalEvent} records to an append-only JSONL file on
state transitions of interest. Critical to evidence preservation, the watcher
captures container state \emph{synchronously} on each pod modification event before
any asynchronous query could complete.

\textit{Layer~2---Operational Memory Store.} A SQLite database operating in
WAL mode stores events, causal edges, point-in-time snapshots, and pattern
definitions. SQLite is chosen deliberately for its suitability to this workload:
the collector produces a single-writer, append-only event stream with no
concurrent write contention, and WAL mode provides non-blocking reads for
concurrent query access. This is not a limitation of the architecture but an
appropriate fit for the workload profile. Production multi-cluster deployments
would replace SQLite with a distributed backend such as PostgreSQL or Parquet
files in object storage; the query interface in Layer~3 is backend-agnostic.

\textit{Layer~3---Query Interface.} Three canonical queries address distinct
operational needs. Q1 (\texttt{causal} chain) reconstructs the causal
predecessors of a given event. Q2 (\texttt{pattern} history) retrieves all
instances of a given causal pattern within a time window. Q3
(\texttt{state-at}) returns the frozen state of an object at a specified
timestamp---even after the object has been deleted from the cluster.

\textit{Layer~4---Integration Surface.} A lightweight HTTP API exposes the
three canonical queries as REST endpoints, enabling integration with AI-assisted
diagnosis systems, alert pipelines, and incident management platforms.

\subsection{Causal Pattern Encoding}

OMA encodes three causal patterns as first-class definitions. Each pattern
specifies a sequence of event types with roles (trigger, precursor, evidence,
effect), temporal windows, and confidence scores for edge construction.
Confidence scores are assigned based on the strength of the domain-specific
temporal constraint: a score of 1.0 is assigned when the relationship is
deterministic given the Kubernetes object model (e.g., OOMKillEvidence for
pod $P$ within 90 seconds of OOMKill for pod $P$ cannot have any other cause),
and 0.9 when the relationship is highly probable but not exclusive (e.g.,
NodeMemoryPressure preceding an OOMKill on the same node may have contributing
causes beyond memory pressure alone).

\textit{Pattern P001 (OOMKill Causal Chain)} encodes the sequence:
NodeMemoryPressure [precursor, 300\,s window, conf=0.9] $\rightarrow$ OOMKill
[trigger] $\rightarrow$ OOMKillEvidence [evidence, 90\,s window, conf=1.0]
$\rightarrow$ ContainerTerminated [effect, 10\,s window]. The 90-second window
for OOMKillEvidence directly encodes the evidence horizon.

\textit{Pattern P002 (ConfigMap Env Var Silent Misconfiguration)} encodes:
ConfigMapChanged [trigger] $\rightarrow$ PodNotRestarted [absence signal,
120\,s window]. The absence of a pod restart following a ConfigMap change is
itself a diagnostic signal when pods consume the ConfigMap as environment
variables. OMA captures the ConfigMap content hash delta at the moment of
change.

\textit{Pattern P003 (ConfigMap Volume Mount Symlink Swap)} encodes:
ConfigMapChanged [trigger] $\rightarrow$ KubeletSync [propagation, 90\,s
window]. The kubelet's atomic symlink swap propagation window is encoded as
a temporal range, enabling OMA to measure actual propagation latency per update.

\subsection{Causal Edge Construction}

Causal edges are constructed automatically by the ingestion layer as events are
loaded into the memory store. For each incoming OOMKillEvidence event, the
ingest layer queries for OOMKill events on the same pod within the preceding
90 seconds and constructs a directed edge with confidence 1.0. The temporal
constraint is implemented as a SQL range query on indexed timestamp columns,
with timezone normalization applied to handle Kubernetes timestamps that
include UTC offset suffixes.

\subsection{Fault Tolerance}

The collector uses an at-least-once delivery model: events are written to the
JSONL file synchronously before acknowledgment, and the ingest layer uses
\texttt{INSERT OR IGNORE} semantics with deterministic event IDs for idempotent
processing. If the collector crashes mid-run, restarting it and re-running
ingest on the accumulated JSONL file produces a correct database with no
duplicate edges. This model accepts the theoretical possibility of duplicate
event delivery in exchange for guaranteed event preservation--the correct
trade-off for evidence preservation under the evidence horizon constraint.

\section{Implementation}
\label{sec:implementation}

\subsection{Go Collector}

The collector is implemented in Go~1.21 using the \texttt{k8s.io/client-go}
library. It comprises three concurrent watchers --NodeWatcher, PodWatcher, and
ConfigMapWatcher---each running in a separate goroutine with context-based
cancellation. The PodWatcher inspects \texttt{ContainerStatus} on every pod
modification event, checking both \texttt{State.Terminated} and
\texttt{LastTerminationState.Terminated} for each container. When an OOMKill
is detected, the watcher immediately captures container resource limits and
requests, ConfigMap and Secret references from the pod spec, the pod's QoS
class, restart count, and a node state snapshot obtained synchronously from
the NodeWatcher cache. The synchronous capture is critical: by the time an
asynchronous query would complete, the pod may have restarted, and the
termination state overwritten.

\subsection{Storage Layer}

The storage layer is implemented in Python~3.11 with no dependencies beyond
the standard library. The \texttt{ingest.py} script reads JSONL files with
streaming semantics, inserting events into SQLite with \texttt{INSERT OR IGNORE}
for idempotency and constructing causal edges on each insertion. Timestamps
from the Kubernetes API include UTC offset suffixes (e.g., \texttt{-05:00});
the ingest layer normalizes these to bare ISO~8601 strings before SQLite
comparison to avoid \texttt{strftime} failures on non-UTC offsets.

\subsection{Schema Design}

The SQLite schema comprises four tables. The \texttt{event} table stores
immutable event records with indexed columns for timestamp, event type,
pattern ID, pod name, namespace, and node name. The \texttt{causal\_edges}
table stores directed edges with foreign key references, pattern ID, confidence
score, and edge type. The \texttt{snapshot} table stores serialized object
state as JSON blobs with composite indexes on
(\texttt{object\_kind}, \texttt{object\_name}, \texttt{timestamp}).
WAL mode is enabled for concurrent reader support without blocking the writer.

\section{Evaluation}
\label{sec:evaluation}

\subsection{Experimental Setup}

We evaluate OMA through three scenario validations, a 30-run statistical
latency analysis, and a concurrent stress evaluation. All experiments were performed  in two independent environments. Environment~1 is a local Minikube cluster
(Kubernetes~1.31) with 3 nodes on Apple M-series hardware. Environment~2 is
an Azure Kubernetes Service cluster (version~1.32.10) with 2~Standard\_B2s
nodes (2\,vCPU, 4\,GB RAM) in \texttt{eastus}. All raw JSONL files, SQLite
databases, query outputs, and automation scripts are committed to the public
repository under \texttt{docs/poc-results/} for independent verification.

\begin{figure}[t]
  \centering
  \includegraphics[width=\columnwidth]{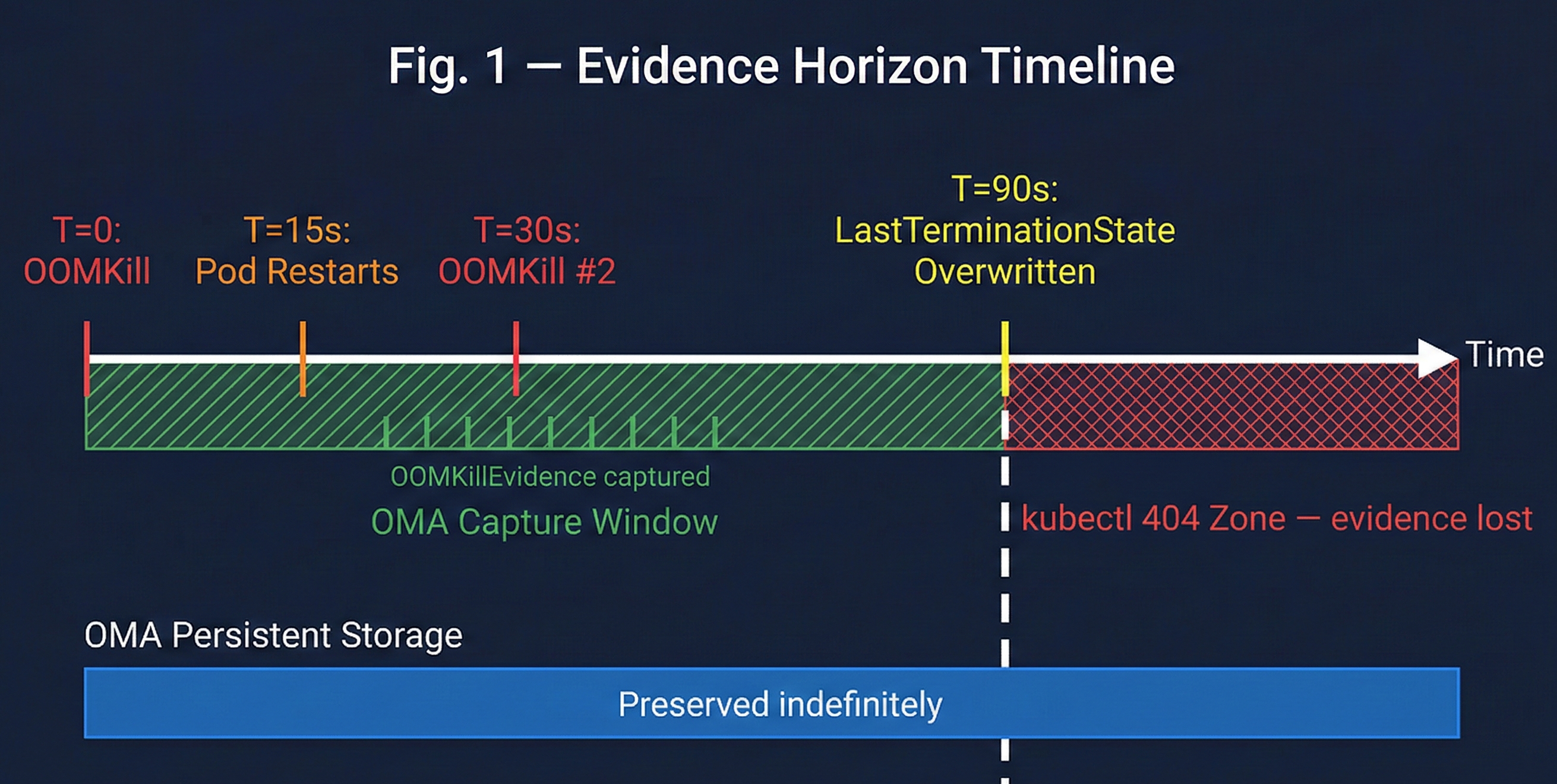}
\caption{The 90-second evidence horizon in Kubernetes.
\texttt{LastTerminationState} is overwritten on each container restart
(T=0\,s: OOMKill; T$\approx$15\,s: first restart; T$\approx$90\,s:
evidence overwritten), permanently discarding OOMKill forensic data
including exit code, resource limits, and ConfigMap references.
OMA captures and preserves all evidence synchronously before rotation
occurs. Beyond the evidence horizon, \texttt{kubectl describe} returns
incomplete data and deleted pods return HTTP~404; OMA point-in-time
snapshots remain queryable indefinitely.}
  \label{fig1}
\end{figure}

\subsection{Scenario Validation (P001, P002, P003)}

Table~\ref{tab1} summarizes the scenario validation results in both environments.
For P001 on Minikube, the collector captured 30--31 events and 13 causal edges
in two independent runs. In AKS, 20 events and 8 edges were captured with
exit code 137 (SIGKILL), confirming correct propagation of the OOM signal through the
AKS container runtime. The Q3 state-at query returned the complete frozen pod
state after deletion, confirming evidence preservation beyond the evidence
horizon.

For P002, both Minikube and AKS pods reported \texttt{FEATURE\_FLAG=disabled}
after ConfigMap was patched to \texttt{enable}, with zero restarts. OMA
captured the hash delta of the content and the list of changed keys, providing an
auditable record invisible to \texttt{kubectl} and Prometheus.

For P003, ConfigMap volume mount propagation completed within 30 seconds on
AKS, confirmed by the pod log detecting the atomic symlink swap. OMA timestamped
both the ConfigMap change and the propagation completion.

\begin{table}[!t]
\caption{Scenario Validation Results}
\label{tab1}
\centering
\setlength{\tabcolsep}{5pt}
\begin{tabular}{|p{0.25\linewidth}|p{0.18\linewidth}|p{0.18\linewidth}|p{0.22\linewidth}|}
\hline
\textbf{Metric} & \textbf{Minikube Run~1} & \textbf{Minikube Run~2} & \textbf{AKS 1.32.10} \\
\hline
Total events (P001)   & 30    & 31    & 20   \\
OOMKill events        & 6     & 6     & 4    \\
OOMKillEvidence       & 16    & 16    & 10   \\
Causal edges          & 13    & 13    & 8    \\
Snapshots             & 1     & 1     & 1    \\
Exit code             & 137   & 137   & 137  \\
Edge confidence       & 1.0   & 1.0   & 1.0  \\
P002 captured         & Yes   & —$^{a}$ & Yes  \\
P003 captured         & Yes   & —$^{a}$ & Yes  \\
\hline
\multicolumn{4}{|p{0.88\linewidth}|}{\footnotesize $^{a}$Run~2 repeated
P001 only for statistical latency analysis; P002 and P003 were
validated in Run~1 (Minikube) and independently on AKS.} \\
\hline
\end{tabular}
\end{table}

\subsection{Statistical Latency Analysis (30 Runs)}

To characterize causal edge construction latency with statistical
confidence, we conducted 30 independent runs of the P001 OOMKill
scenario on Minikube, each run comprising a clean namespace creation,
collector start, pod deployment, 75-second observation window, and
causal edge construction. In total, 242 causal edges were analyzed
across the 30 runs.

The latency distribution is bimodal, reflecting two structurally distinct
edge types. \emph{Intra-cycle edges} link an OOMKillEvidence event to the
OOMKill event from the same restart cycle---the evidence is captured within
milliseconds of the kill. \emph{Cross-cycle edges} link a later OOMKillEvidence
event back to an earlier OOMKill event from a previous restart cycle, where
the latency reflects the actual restart interval rather than processing delay.
Table~\ref{tab2} reports both distributions separately.

\begin{table}[!t]
\centering
\caption{Causal Edge Construction Latency (30 Runs, 242 Edges)}
\label{tab2}
\setlength{\tabcolsep}{5pt}
\begin{tabular}{|p{0.30\linewidth}|p{0.10\linewidth}|p{0.12\linewidth}|p{0.12\linewidth}|p{0.12\linewidth}|}
\hline
\textbf{Edge Class} & \textbf{Count} & \textbf{Min} & \textbf{Mean} & \textbf{Max} \\
\hline
Intra-cycle ($<$100\,ms) & 88  & 0.089\,ms & 0.702\,ms  & 2.607\,ms  \\
Cross-cycle ($\geq$100\,ms) & 154 & 903\,ms & 12,708\,ms & 31,454\,ms \\
\hline
\multicolumn{5}{|p{0.88\linewidth}|}{Intra-cycle edges confirm sub-millisecond
synchronous evidence capture within the same restart cycle. Cross-cycle
latency reflects restart interval timing, not processing delay.} \\
\hline
\end{tabular}
\end{table}

The intra-cycle mean of 0.702\,ms with a range of 0.089--2.607\,ms across
88 observations demonstrates that OMA's synchronous capture model successfully
preserves evidence well within the evidence horizon on every run. The
cross-cycle distribution (mean 12,708\,ms) reflects the CrashLoopBackOff
backoff timing between restart cycles---this is expected behavior demonstrating
that OMA correctly links evidence events to their originating OOMKill across
multiple restart cycles within the 90-second window.

\subsection{Stress Evaluation}

To assess OMA's behavior under concurrent load, we deployed 5, 10, and 20
simultaneous crash-looping pods (each with a 64\,Mi memory limit and 128\,Mi
allocation) on a single Minikube cluster and ran each level for 120 seconds.
Table~\ref{tab3} reports the results.

\begin{table}[!t]
\centering
\caption{Stress Evaluation: Concurrent OOMKill Pods}
\label{tab3}
\setlength{\tabcolsep}{5pt}
\begin{tabular}{|p{0.10\linewidth}|p{0.14\linewidth}|p{0.16\linewidth}|p{0.14\linewidth}|p{0.14\linewidth}|}
\hline
\textbf{Pods} & \textbf{Events} & \textbf{Events/sec} & \textbf{Edges} & \textbf{RAM (MB)} \\
\hline
5  & 95  & 0.77 & 51  & 7.9 \\
10 & 175 & 1.43 & 90  & 8.2 \\
20 & 355 & 2.86 & 197 & 8.8 \\
\hline
\end{tabular}
\end{table}

Event ingestion scales linearly with pod count: doubling pods from 5 to 10
doubles the event rate (0.77 to 1.43 events/sec), and doubling again to 20
doubles again (2.86 events/sec). Collector memory consumption remains flat
at 7.9--8.8\,MB across all three load levels, confirming that the streaming
JSONL model accumulates no in-memory state proportional to event volume.
CPU utilization remained below the measurement threshold ($<$0.1\%) throughout,
consistent with the I/O-bound nature of the collector workload.

\subsection{Comparison with Existing Tools}

Table~\ref{tab4} compares OMA against \texttt{kubectl} and Prometheus across
key diagnostic capabilities.

\begin{table}[!t]
\centering
\caption{Capability Comparison}
\label{tab4}
\setlength{\tabcolsep}{5pt}
\begin{tabular}{|p{0.40\linewidth}|p{0.13\linewidth}|p{0.16\linewidth}|p{0.13\linewidth}|}
\hline
\textbf{Capability} & \textbf{kubectl} & \textbf{Prometheus} & \textbf{OMA} \\
\hline
OOMKill evidence after restart  & $<$90\,s & No        & Indefinite \\
Resource limits at kill time    & $<$90\,s & Approx.   & Exact, frozen \\
ConfigMap in effect at failure  & No       & No        & Refs + hash \\
Stale env var detection         & No       & No        & Yes (P002) \\
ConfigMap propagation latency   & No       & No        & Yes (P003) \\
State of deleted objects        & 404      & Partial   & Yes (Q3) \\
Causal edges between events     & No       & No        & Yes \\
Pattern recurrence detection    & No       & Via rules & Yes (Q2) \\
\hline
\end{tabular}
\end{table}

\begin{figure}[t]
  \centering
  \includegraphics[width=\columnwidth]{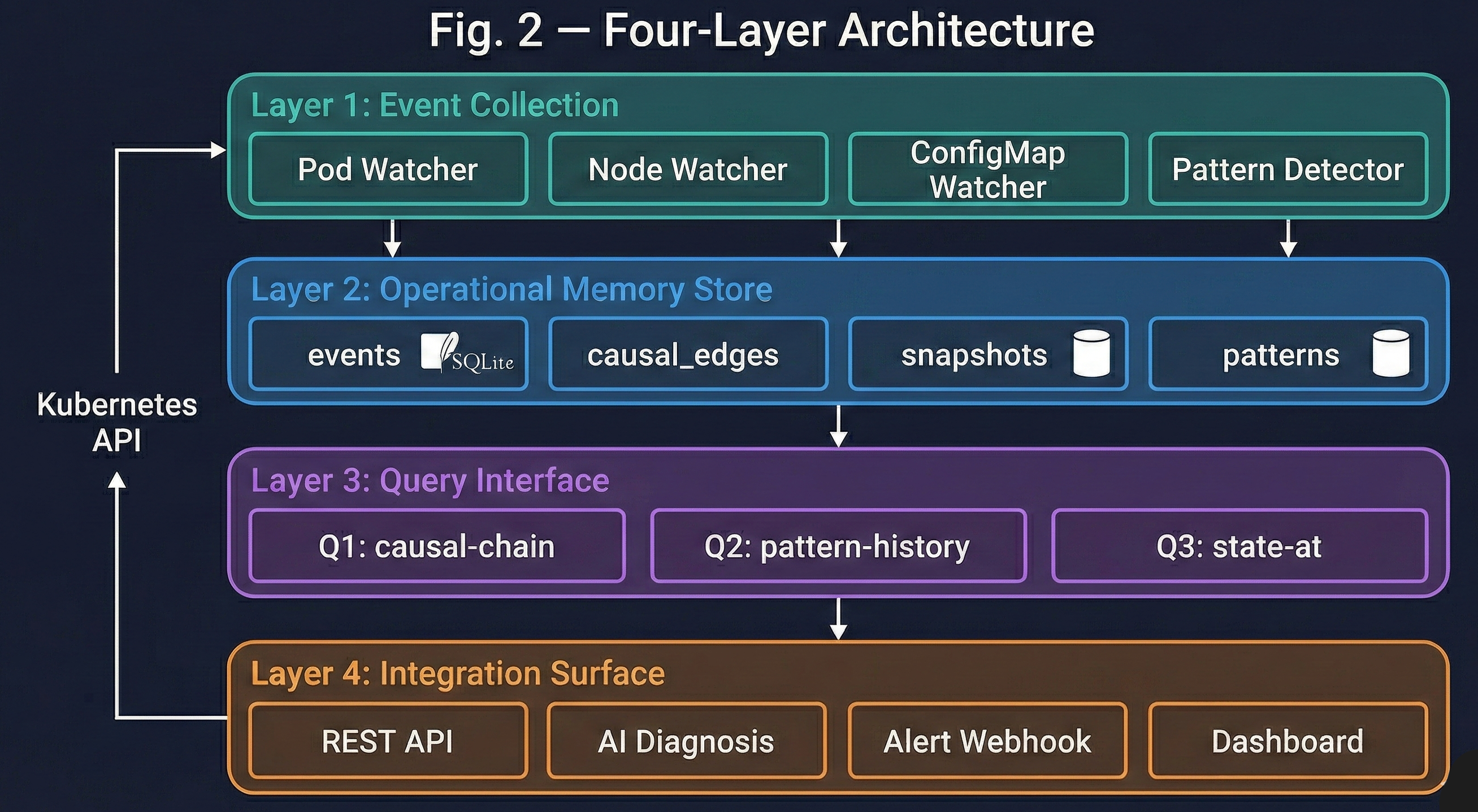}
  \caption{OMA four-layer architecture. Layer~1 (Go collector) subscribes
  to Kubernetes API watch streams via \texttt{client-go} informers and emits
  structured \texttt{CausalEvent} records to an append-only JSONL buffer.
  Layer~2 (SQLite WAL-mode memory store) ingests events and constructs causal
  edges automatically on insertion. Layer~3 (query interface) exposes three
  canonical queries: Q1 causal-chain, Q2 pattern-history, Q3 state-at.
  Layer~4 (integration surface) provides a REST API for downstream consumers.}
  \label{fig2}
\end{figure}

\begin{figure}[t]
  \centering
  \includegraphics[width=\columnwidth]{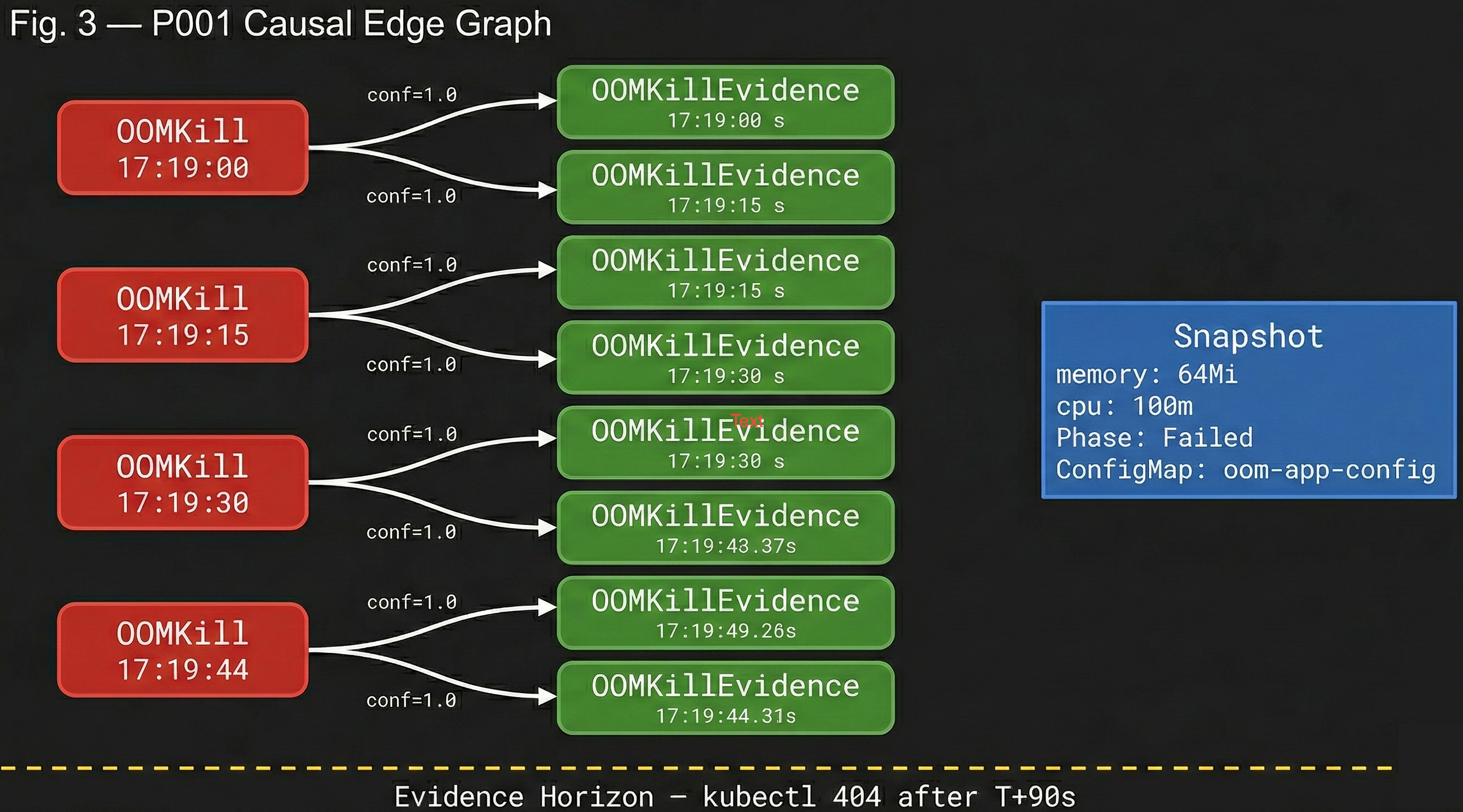}
    \caption{P001 causal edge graph from the AKS run
    (\texttt{aks-nodepool1-78296979-vmss000000}). Four OOMKill events (red)
    are linked to eight OOMKillEvidence events (green) via directed causal
    edges with confidence 1.0. Intra-cycle edge construction latency ranges
    from 0.089\,ms to 2.607\,ms (mean 0.702\,ms across 30 runs);
    cross-cycle edges reflect restart interval timing (900\,ms--31{,}454\,ms).
    A point-in-time snapshot (blue) preserves the complete pod
    state---memory limit 64\,Mi, CPU 100\,m, Phase Failed, ConfigMap
    \texttt{oom-app-config}---after pod deletion. The evidence horizon
    marker indicates the point beyond which \texttt{kubectl} returns
    HTTP~404 and native Kubernetes state is irrecoverable.}
  \label{fig3}
\end{figure}

\section{Discussion}
\label{sec:discussion}

\subsection{Limitations}

The NodeMemoryPressure $\rightarrow$ OOMKill causal edge was not observed in
our experiments. The AKS Standard\_B2s nodes have 4\,GB RAM; killing a single
64\,Mi pod does not create sufficient node-level memory pressure to trigger the
NodeMemoryPressure condition. The pattern encoder is implemented and the edge
type is defined; the scenario requires genuine node-level memory exhaustion
across multiple pods and is deferred to future work.

The current collector is namespace-scoped. Production deployment requires either
cluster-wide RBAC permissions or per-namespace collector instances. A Helm chart
for cluster-wide deployment is planned. The SQLite backend is appropriate for
single-cluster workloads; multi-cluster deployments would benefit from a
distributed backend with query federation.

The 30-run statistical evaluation was conducted on Minikube running on Apple
M-series hardware. While this provides statistical confidence in the latency
measurements, production AKS clusters may exhibit different timing
characteristics due to differences in container runtime (containerd vs. Docker),
node hardware, and kubelet configuration. The AKS single-run results (8 edges,
exit code 137) are consistent with the Minikube distribution, but a full
statistical evaluation on AKS is deferred to future work.

\subsection{RBAC and Security Considerations}

The collector requires read access to Pod, Node, and ConfigMap objects in the
watched namespace. A minimal RBAC configuration requires a \texttt{ClusterRole}
with \texttt{get}, \texttt{list}, and \texttt{watch} verbs on these resource
types, bound to a dedicated service account. ConfigMap watch access should be
considered carefully in environments where ConfigMaps store sensitive
configuration; the collector captures only metadata and content hashes by
default, not raw ConfigMap values.

\subsection{Production Deployment Considerations}

In production environments with strict compliance requirements, OMA's immutable
JSONL event log constitutes an auditable record of configuration state changes
and container failure reasons. The content hash delta captured for each ConfigMap
change provides a lightweight configuration drift detection mechanism. Resource
overhead is minimal: the Go collector consumes approximately 8--9\,MB RAM and
negligible CPU even under 20 concurrent crash-looping pods, as confirmed by the
stress evaluation.

\section{Conclusion}
\label{sec:conclusion}

We have presented the Operational Memory Architecture, formalized as a
first-class architectural primitive for Kubernetes operational state preservation.
OMA explicitly encodes evidence rotation constraints and causal preservation
as design requirements---an aspect not addressed by existing observability or
logging models. Through scenario validation on Minikube and AKS~1.32.10, a
30-run statistical latency analysis (242 edges, intra-cycle mean 0.702\,ms),
and a concurrent stress evaluation confirming linear event scaling at flat
8--9\,MB RAM, we demonstrate that OMA successfully preserves causal context
that existing tools discard within seconds of occurrence.

The three causal patterns encoded in this work---P001, P002, and P003---represent
a starting point for a broader pattern library. The evidence horizon is a
fundamental property of the Kubernetes architecture; operational memory is a
necessary complement to metrics, logs, and traces in achieving full-stack
observability. We release the complete implementation, experimental data, and
evaluation scripts as open-source software and invite contributions of
additional pattern encoders and storage backends.

\appendices

\section*{Acknowledgments}
The author thanks the open-source Kubernetes and Go communities whose
tooling made this work possible.

\end{document}